# Inventions on using colors in Graphical User Interfaces
## A TRIZ based analysis

**Umakant Mishra**

Bangalore, India

http://umakantm.blogspot.in

**Contents**

## 1. Introduction

Color is an important aspect of any graphical user interface. Color is used to improve various aspects of the GUI, such as:

⇛ Making the GUI more attractive,
⇛ Differentiating one item or group of items from the other. For example, showing currently selected item or window in a different color.
⇛ Assigning weights to components by using intensity of color.
⇛ Soothing to different moods, personalities etc.

However, there are difficulties in using colors too. Improper use of color can give adverse effects. Wrong colors at wrong place make the GUI look clumsy and confusing. Besides the aesthetics issues there are many other issues involved with colors. For example;

⇛ Displaying and selection of as many as 16 million colors.
⇛ Simplifying color customization of GUI according to user choice.
⇛ Managing color compatibility among different GUI products.
⇛ Color conversion for incompatible environments, and so on.

For example, it is desirable that the color scheme of a GUI is customizable by the user. But it may have adverse effects, such as; the user may have to alter the color schemes in every computer he uses. Again the altered color scheme may be confusing for another person who wishes to use that system. This situation creates the following contradiction.

**"The color of the GUI should be customizable to suit user preference. But at the same time it should not be customizable, as that will cause annoyance and confusion to others."**

Taking another example, the user should be able to precisely select from all 16 million colors. But displaying all 16 million colors will be confusing to the user. This creates the following contradiction.

**"The user should be displayed all 16 million colors to select the desired color precisely, and the user should not be displayed all 16 million colors as that would create confusion and difficulty in selection."**

There are many more contradictions relating to use of color in GUI. Let's discuss some of the inventions on GUI color selection and analyze how they resolve the existing contradictions to solve the problem.

## 2. Inventions on GUI Color Selection

### 2.1 Apparatus and method for color selection (Patent 5103407)

**Background problem**

There is a need to display all 16 million colors and permit the user to choose from among the 16 million colors including their hue, saturation and luminance. The prior art (Apple Mac-II) has disposed a two-dimensional color circle with varying hue and saturation levels. The user chooses a color from a palette of 256 colors and then corrects it by choosing a luminance level from the luminance bar and hue and saturation levels from the color circle. Upon choosing a luminance level the system displays the resultant color. This method of color selection is slow.

**Solution provided by the invention**

Gabor disclosed a new method of color selection (US Patent 5103407, Assigned by Scitex Corporation, issued in Apr 1992), which is relatively quick and efficient. The method contains a first area that displays colors of different hues, and a second a area which displays a color in different saturation and luminance. The user refines his desired color by selecting from both these selection areas.

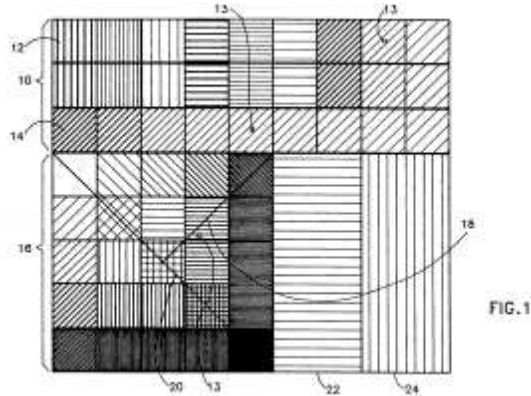

FIG.1

The user selection apparatus also has an area to display a reference color (which has been selected from an image on the screen) and a selected color, which is supposed to replace the reference color.

**TRIZ based analysis**

As all the colors are difficult to put in a single selection box, the invention uses two selection areas, the first displays colors of different hue (in predetermined luminance and saturation) and the second displays (the selected color in) different luminance and different saturation (Principle-1: Segmentation).

The selection uses grids to display color schemes. The first area comprises a plurality of one-dimensional grids, the second area displays colors in a two-dimensional grid (Principle-1: Segmentation).

**2.2 Pattern and color abstraction in a graphical user interface (Patent 6239795)**

**Background problem**

There are various elements of a graphic user interface such as windows, menus, scroll bars, toolbox etc. Each application developer can define his own nonstandard controls and window types as desired. Consequently there may be three applications running on a desktop each having windows of different look and feel. This dissimilarity in appearance and behavior between applications can be annoying and confusing to a user.

**Solution provided by the invention**

Ulrich et al., provided a method of increased flexibility and control over the appearance and behavior of a user interface (Patent 6239795, assigned by Apple Computer Inc., issued May 2001). According to the invention, sets of objects are grouped into themes to provide a user with a distinct overall impression of the interface. The user can switch between the themes, even at runtime, to change the desired appearance and behavior.

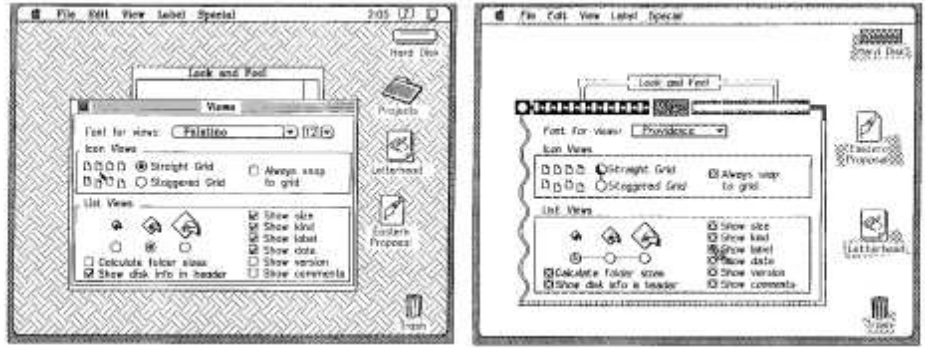

(This invention is later continued in Patent No 6466228, by the same inventors, issued in Oct 2002).

**TRIZ based analysis**

The invention groups the appearance and behavior of objects into themes. (Principle-5: Merging).

The user chooses themes to change the appearance and behavior of the graphic user interface even during runtime (Principle-15: Dynamize).

## 2.3 Converting TV unsafe colors to TV safe colors (Patent 6466274)

**Background problem**

The convergence of computers and televisions makes it desirable to be able to display graphical color information created for a computer on a television. There are different methods used in the industry to convert computer colors to TV safe colors.

One method is to simply avoid the use of TV unsafe colors altogether. But this has the effect of eliminating 25 percent of colors in the source color palette.

Another method is to forcibly convert the TV unsafe colors in the source palette to TV safe colors (Principle[16]:"Partial or Excessive Action"). But this destroys the color trends and interrelationships of the original palette.

**Solution provided by the invention**

Brian White invented a method (US patent 6466274, assigned to Corporate Media Partners, Oct 2002) of transforming a color palette with TV unsafe colors into a palette that contains with only TV safe colors where the transformed color palette is nearly identical when perceived by the human eye. This is achieved by using a conversion table mapping the TV unsafe colors to TV safe colors.

## TRIZ based analysis

The invention converts the TV unsafe colors to its nearest identical TV safe color (Principle-16: Partial or excessive action, Principle-36: Conversion).

The conversion is achieved by using a conversion table mapping the TV unsafe colors to TV safe colors. (Principle-10: Prior action).

## 2.4 Contrasting graphical user interface pointer (Patent 6486894)

**Background problem**

The GUI is mostly operated by a pointing device like a mouse or trackball. Most GUI systems use color display screens either predefined or definable by the user. The pointers in these color systems may be colored to get the aesthetic and functional effect based on the background and foreground colors. The customizable and programmable color selection in the GUI systems gives rise to problems of cursor/ pointer visibility. If the pointer is displayed in the same or similar color as a color used in the background or foreground color, it may be difficult to see or sometimes become invisible.

The problem also persists in monochrome or gray-scale display systems when the shades used to mimic colors are similar to that of the pointer. This problem of cursor visibility reduces efficiency and causes great concern for visually challenged users.

**Solution provided by the invention**

Patent 6486894 (invented by Abdelhadi et al., assigned to IBM Corporation, issued Nov 2002) discloses an improved method of displaying mouse pointer in a GUI system using color or monochrome display screens.

The invention analyses the background and foreground colors over which the pointer is to appear and ensures that the pointer colors are always contrasting to the background and foreground colors over which the pointer resides.

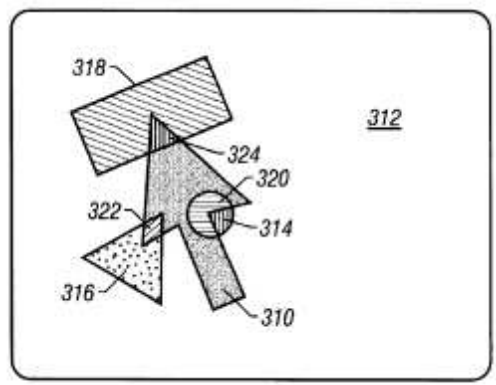

TRIZ based analysis]

The invention continuously evaluates the color of the existing screen at the location the pointer is to appear and computes a contrasting color on a point-by-point basis (Principle-20: Continuous action).

The pointer is displayed in a color or shade that contrasts the background and foreground color for high visibility (Principle-32: Color change).

## 2.5 Half-toning without a full range of equally-spaced colors (Patent 6580434)

### Background problem

Most graphics today use true color or high color using as many as 16 million colors. But there are many machines that do not support more than 16 colors or 256 colors. How to adjust the display of high color graphics to the few colors available in the display?

Out of different methods available, one is half-toning. Half-toning displays or prints an image with limited number of color levels. The intensity of each pixel is calculated from a halftone matrix.

The conventional half-toning does not allow using a subset of half intensity colors. So only $n^3$ colors (8 or 27 and so on) can be used. As a result the users having 16-color display get only the benefits of eight colors and not all 16.

### Solution provided by the invention

Donald Curtis invented a new half-toning mechanism (US Patent 6580434, assigned by Microsoft, June 2003) that uses all of the colors available from the display. The halftone value is chosen from a half-tone matrix using a different logic that allows using the half intensity colors if the RGB value of a pixel falls less than a predetermined value.

### TRIZ based analysis

The invention uses a half-tone matrix for color conversion (Principle-24: Intermediary).

The half-tone matrix is calculated using a different logic that allows half intensity colors if the RGB value falls less than a predetermined value (Principle-16: Partial or excessive action).